# Using Stochastic local search in designing microstrip antenna for 5G communication

Sunit Shantanu Digamber Fulari, Department of Electronics and Communication, Chandigarh University, Assistant professor, Govt college of Arts, Science and Commerce, Khandola, Marcela, Goa

**Abstract:** Well this paper defines methods to explore what is known as the local search problem, this local search is what we are going to use in antenna to antenna and antenna to device communication. The local search algorithm searches for best next for search this is in turn used by us in antenna pairing. This is prominently known as stochastic local search, We are going to design 5G microstrip antenna operating between 2.4GHz to 24 Ghz of operation. This speaks of a very novel idea which though was used in late sixties when microstrip was in operation but the idea is having potential.

**I. Introduction:** We use the stochastic local search in our proposed work on microstrip 5G antenna. This is a old method involving the major travelling salesman problem which involved the shortest path taken to deliver. Our antenna system and elements should find the shortest distance connecting with each other. Heuristic search problem is finding a better way each time. Antenna radiate energy with the help of feed getting reflected on the reflector and connect with prominent next antenna and potential mobile devices. We are trying to make this radiation effective in a smart way so that the shortest distance is covered by the antenna connecting with other antennas or devices. This is a very potent tricky problem to be solved as we are trying to solve the problem of radiation being extradiated into space which we try to minimize, we already know this is harmful to life in general. Specific aborption rate is a term associated with the mobile devices high frequency is related to high SAR. Antenna communication is a very complex subject we try to use the stochastic local search in designing and implementing our antenna setup. In the figures below we have used the smallest distance connecting the antennas.

**Figure I: Shortest distance between antennas**

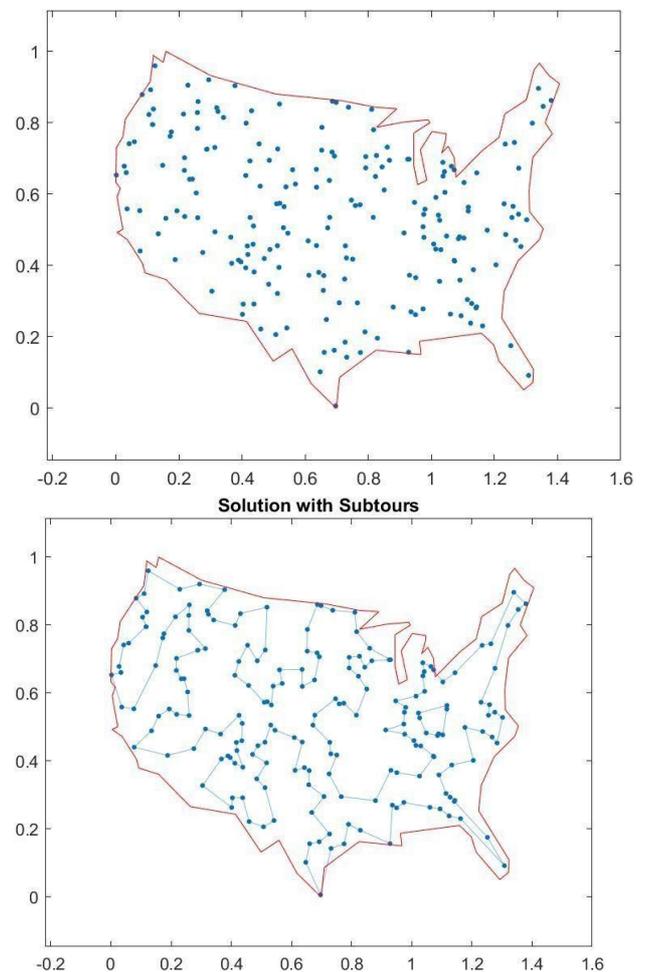

The Antenna towers will communicate with surrounding systems with its radiation being given out, this way it will use the smallest distance smartly to connect with each other.

**Figure II: Antenna towers with potential matrix calculation.**

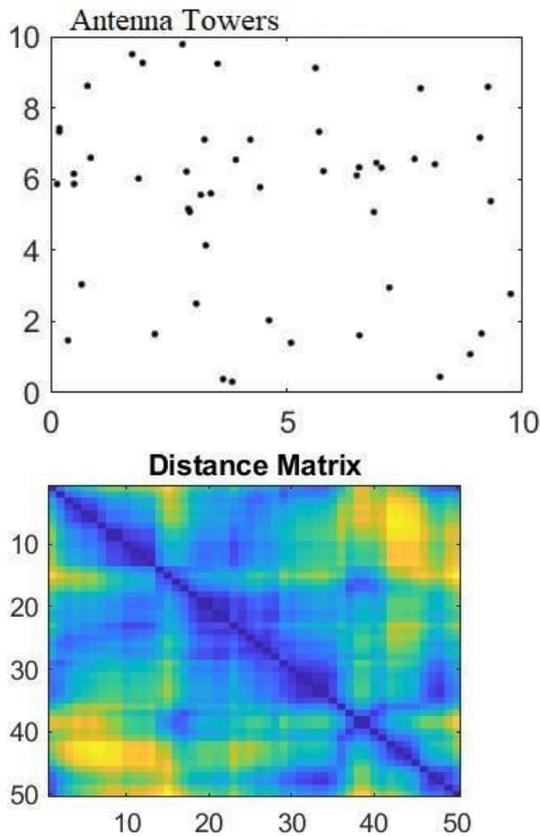

The distance matrix is used to find the shortest distance smartly.

**Figure III: Shortest Linear distance between elemental system**

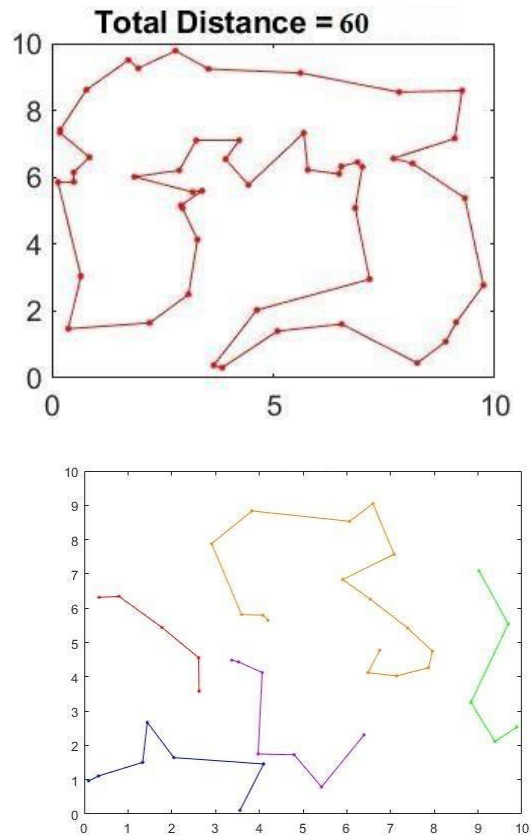

The above figure is on the smart connection between the antennas to form the overall profile of the antenna system.

This is a very important concept as antenna radiation transmission also causes lot of pollution, which we try to minimize through our research. We use the shortest distance between the systems. Radiowaves are tried to minimize through our design.

## II. Antenna as the ears and eyes

Antennas were founded to improve the communication by exchange of messages. In the prior foundation of antennas they have evolved today into various shapes and sizes due to the applications they adhere to. Horn shaped antennas are well known to be used in space communication. In our previous paper we have elaborated on the useof horn shaped antenna. This paper will be on more discrete analysis of more antenna elements. The various shapes of antenna makes them viable to be used in various applications. We are exploring mobile antenna from 2.4Ghz to 24 Ghz in time domain analysis. We are going to study micropatch antenna for mobile communication antenna. Thickness is 1.5mm.

The major equations used in the design of microstrip edge extended antenna.

Width is given by:

$$W = \frac{c}{2f\sqrt{\frac{\varepsilon+1}{2}}}$$

C=velocity of light    W=36.27mm.

$$\varepsilon_{reff} = \frac{\varepsilon_r+1}{2} + \frac{\varepsilon_r-1}{2}\left[1+12\frac{h}{w}\right]^{-\frac{1}{2}}$$

$$L_{eFF} = \frac{c}{2f_v\sqrt{\varepsilon r_{e_F}}}$$

$$\Delta L = 0 \cdot 412h \frac{(\varepsilon_{r_{eFF}}+0.3)\left[\frac{W}{h}+0.264\right]}{(\varepsilon_{re_f}-0.25-8)\left(\frac{W}{h}-0.8\right)}$$

$$L = L_{eFF} - 2\Delta L$$

Length is 2W= 72.54 as L=2*W,

These were the basic formulas used in our design of system.

The operating frequency is given by 2.4Ghz, The FR4 has a metallic dielectric constant of $\varepsilon_r$=4.7, height if the dielectric substrate is h=0.035, $L_g$=2*l, $W_g$=2*W, $F_i$=6h/2=4.8mm. length of the feed line, the gap between the patch and the inset fed is usually 1mm, input impedance is usually 50Ohms, the width of the microstrip feed line is Wf,

Figure IV:Design of microstrip(L=2W)

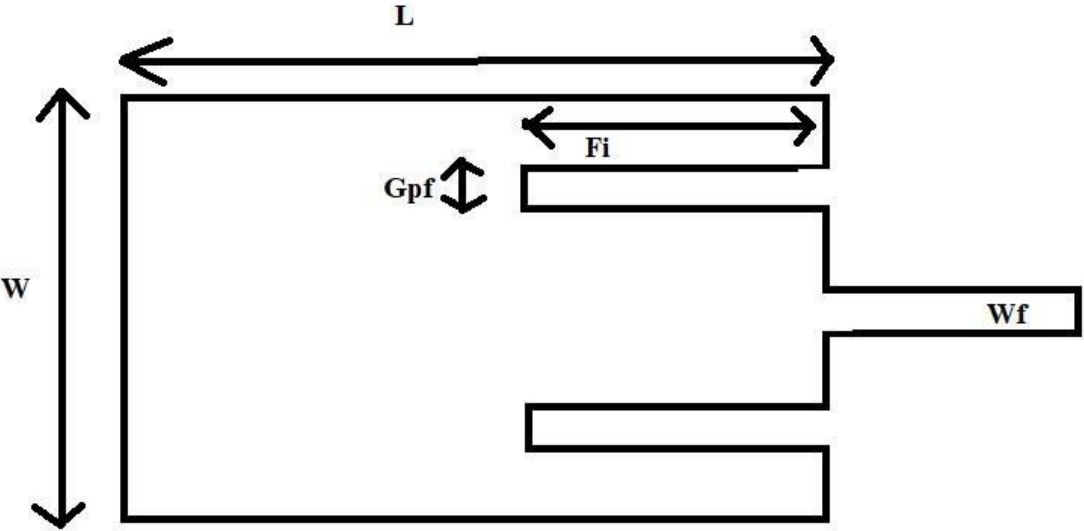

This is a linear approximation of our microstrip patch antenna design. The length is double the width and the other approximations are as follows.

Table

| Parameter | mm |
|---|---|
| W | 36.27 |
| L | 72.54 |
| Fi | 4.8 |
| Wf | 2.932 |
| Gpf | 1 |
| Lg | 2*L |
| Wg | 2*W |
| Ht(copper thickness) | 0.035 |
| Hs | 1.6 |

Thin microstrip is our proposed model, we use copper annealed with FR4-lossy substrate with copper(annealed) as patch. Empty space is created by Nickel which will be visible at the bottom to slice it off.

**Figure V: The design of 5G microstrip antenna with waveguide, substrate and ground is as follows.**

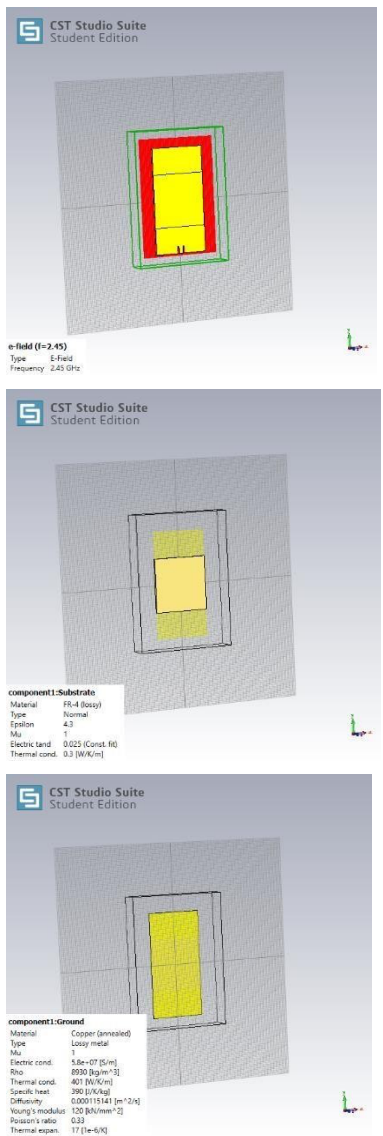

**Figure VI: Received and incident signals in port are as follows.**

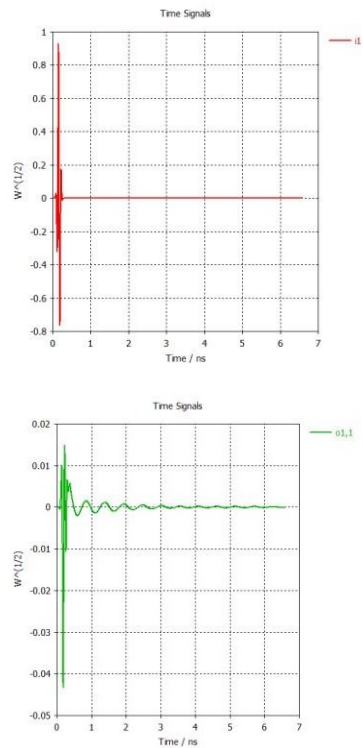

**Figure VII:**
**S parameter as reflection of electromagnetic signal, excitation and impedance of feed line.**

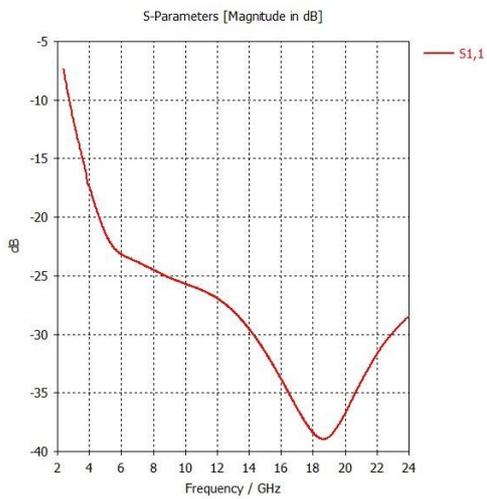

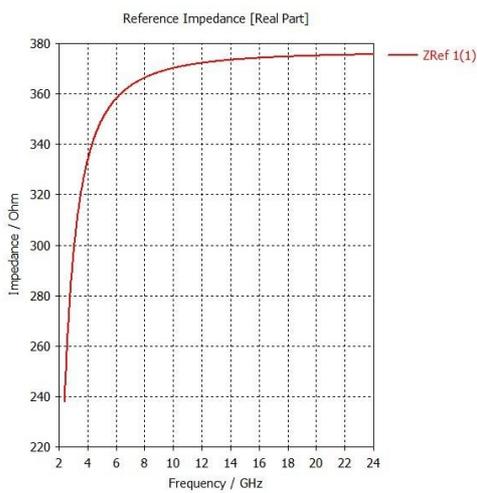

**Figure VIII: How well the stability of the system is known as balance impedance and field excitation magnitude**

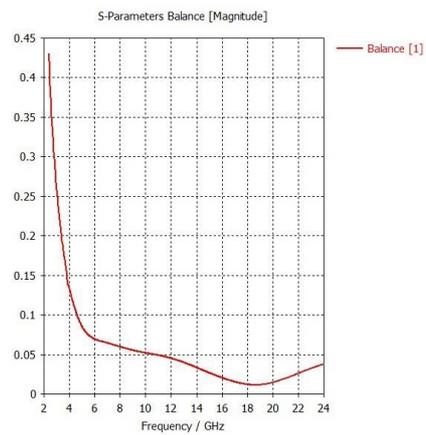

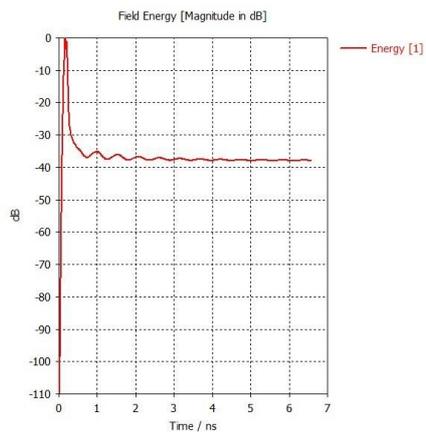

**Figure IX:Excitation power measured in watts**

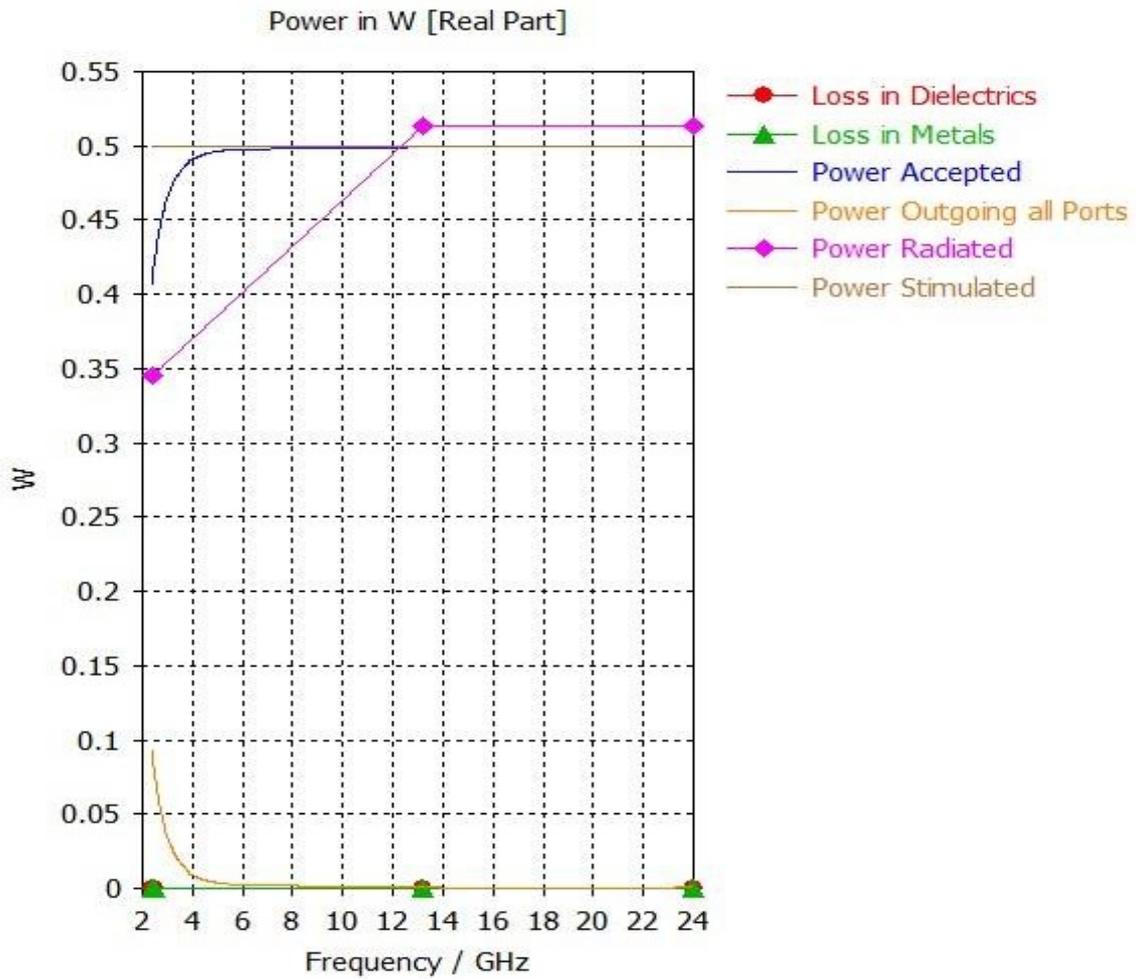

**Figure X:Voltage standing wave ratio, y and z axis**

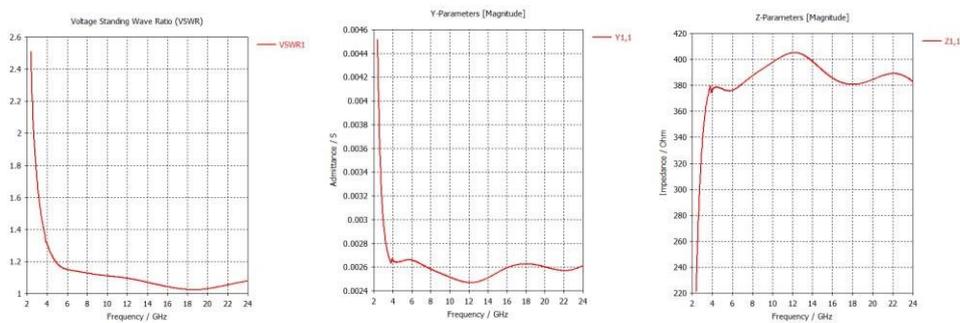

**Figure XI: Surface current, power and power loss density at 2.4Ghz frequency.**

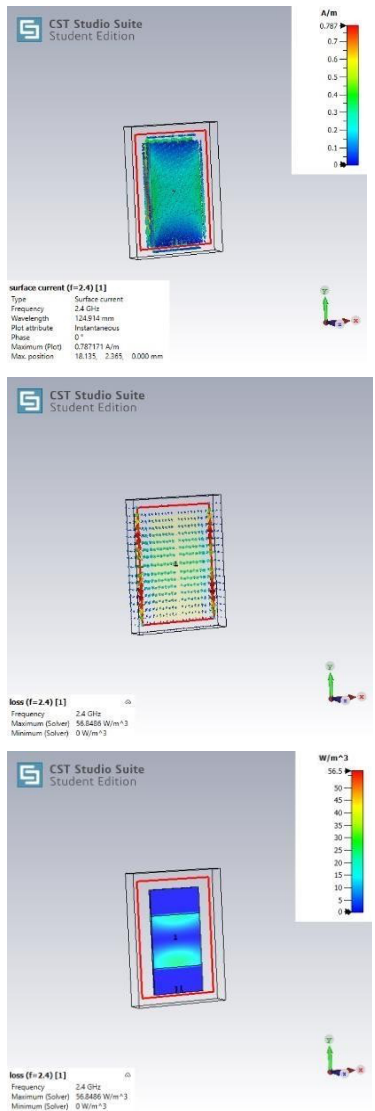

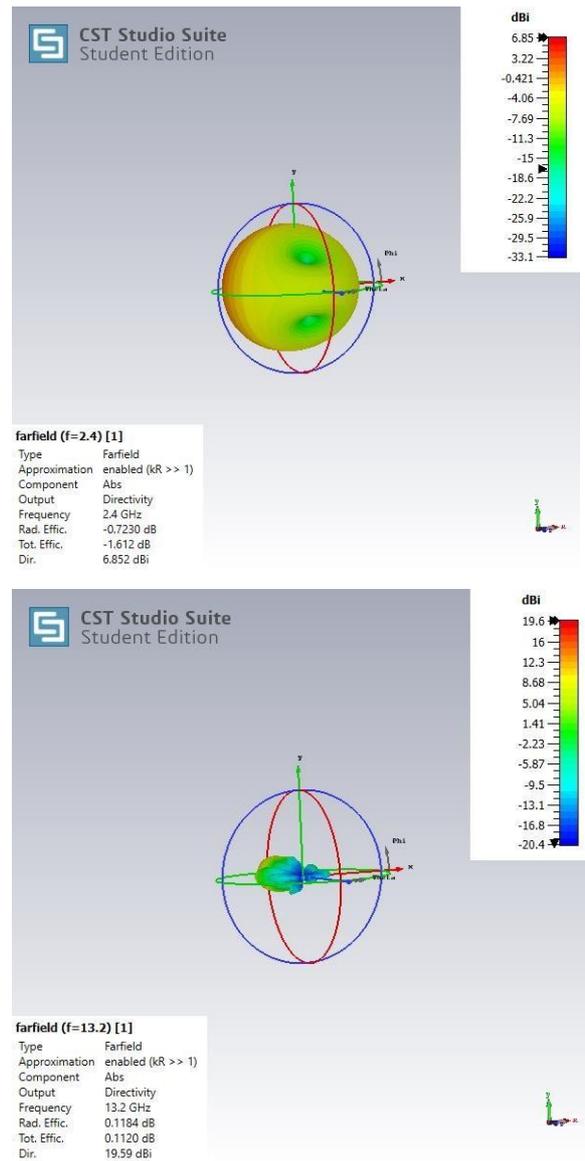

**Figure XII: Far field radiation pattern at 2.4 and 13.2 Ghz.**

**CONCLUSION**: We have simulated the microstrip 5G antenna for 2.4Ghz to 24 Ghz frequency of operation. The radiation pattern shows the effectiveness of the microstrip antenna for enhanced directivity with reduced side lobes. Antenna design is a very discrete task as we want them to operate at minimum power but generating high amounts of power to radiate in all directions which is shown in this design. The transmitted to reflected ratio or VSWR(figure X) is shown gradual decrease to constant value which shows the dominant operation of this antenna.

The stochastic local search is well addressed in this paper which resolves the problem of connectivity between antennas and wireless communication. The local search is a systematic approach of transmission and retransmission of radio waves. The effectiveness of this antenna in prominently transmitting energy in addition to maintaining the connectivity with other antennas and devices makes it a very advantageous system to be built upon.

Declaration:

Funding:No funding received.

Conflict of interest:No conflict of interest.

Availability of data:No data procured.

Code availability:No code available.

# References:

[1] Pozar, D.M., 1992. Microstrip antennas. *Proceedings of the IEEE*, *80*(1), pp.79-91.

[2] Garg, R., Bhartia, P., Bahl, I.J. and Ittipiboon, A., 2001. *Microstrip antenna design handbook*. Artech house.

[3] James, J.R., Hall, P.S. and Wood, C., 1986. *Microstrip antenna: theory and design* (Vol. 12). Iet.

[4] Carver, K. and Mink, J., 1981. Microstrip antenna technology. *IEEE transactions on antennas and propagation*, *29*(1), pp.2-24.

[5] James, J.R., 1989. *Handbook of microstrip antennas* (Vol. 1). IET.

[6] Luk, K.M., Mak, C.L., Chow, Y.L. and Lee, K.F., 1998. Broadband microstrip patch antenna. *Electronics letters*, *34*(15), pp.1442-1443.

[7] Lo, Y.T., Solomon, D. and Richards, W., 1979. Theory and experiment on microstrip antennas. *IEEE transactions on Antennas and Propagation*, *27*(2), pp.137-145.

[8] Pozar, D.M., 1985. Microstrip antenna aperturecoupled to a microstripline. *Electronics letters*, *21*(2), pp.49-50.

[9] Dahele, J.S. and Lee, K.F., 1985, December. Theory and experiment on microstrip antennas with airgaps. In *IEE Proceedings H (Microwaves, Antennas and Propagation)* (Vol. 132, No. 7, pp. 455-460). IET Digital Library.

[10] Nasimuddin, N. ed., 2011. *Microstrip antennas*. BoD–Books on Demand.

[11] Tanaka, M. and Jang, J.H., 2003, June. Wearable microstrip antenna. In *IEEE Antennas and Propagation Society International Symposium. Digest. Held in conjunction with: USNC/CNC/URSI North American Radio Sci. Meeting (Cat. No. 03CH37450)* (Vol. 2, pp. 704-707). IEEE.

[12] Derneryd, A., 1979. Analysis of the microstrip disk antenna element. *IEEE Transactions on Antennas and Propagation*, *27*(5), pp.660-664.

[13] Cengiz, Y. and Tokat, H., 2008. Linear antenna array design with use of genetic, memetic and tabu search optimization algorithms. *Progress In Electromagnetics Research*, *1*, pp.63-72.

[14] Goudos, S.K., Gotsis, K.A., Siakavara,K., Vafiadis, E.E. and Sahalos, J.N., 2013. A multiobjective approach to subarrayed linear antenna arrays design based on memetic differential evolution. *IEEE Transactions on Antennas and Propagation*, *61*(6), pp.3042-3052.

[15] Tseng, L.Y. and Han, T.Y., 2008. Microstrip-fed circular slot antenna for circular polarization. *Microwave and Optical Technology Letters*, *50*(4), pp.1056-1058.

[16] Qu, B.Y., Liang, J.J. and Suganthan, P.N., 2012. Niching particle swarm optimization with local search for multi-modal optimization. *Information Sciences*, *197*, pp.131143.

[17] Capozzoli, A. and D'Elia, G., 2006. Global optimization and antenna synthesis and diagnosis, part two: applications to advanced

reflector antennas synthesis and diagnosis techniques. *Progress In Electromagnetics Research*, *56*, pp.233-261.

[18] Simon, D., Omran, M.G. and Clerc, M., 2014. Linearized biogeography-based optimization with re-initialization and local search. *Information Sciences*, *267*, pp.140-157.

[19] Quijano, J.L.A. and Vecchi, G., 2012. Optimization of a compact frequency-and environment-reconfigurable antenna. *IEEE transactions on antennas and propagation*, *60*(6), pp.2682-2689.

[20] Johnson, J.M. and Rahmat-Samii, Y., 1994, June. Genetic algorithm optimization and its application to antenna design. In *Proceedings of IEEE Antennas and Propagation Society International Symposium and URSI National Radio Science Meeting* (Vol. 1, pp. 326-329). IEEE.

[21] Kim, Y., Keely, S., Ghosh, J. and Ling, H., 2007. Application of artificial neural networks to broadband antenna design based on a parametric frequency model. *IEEE Transactions on Antennas and Propagation*, *55*(3), pp.669-674.

[22] Xu, Z., Li, H., Liu, Q.Z. and Li, J.Y., 2008. Pattern synthesis of conformal antenna array by the hybrid genetic algorithm. *Progress In Electromagnetics Research*, *79*, pp.75-90.

[23] Ding, D. and Wang, G., 2013. Modified multiobjective evolutionary algorithm based on decomposition for antenna design. *IEEE transactions on antennas and propagation*, *61*(10), pp.5301-5307.

[24] Wang, L., Zhang, X. and Zhang, X., 2019. Antenna array design by artificial bee colony algorithm with similarity induced search method. *IEEE Transactions on Magnetics*, *55*(6), pp.1-4.